\documentclass[conference]{IEEEtran}
\usepackage{amsmath, fancybox, epsfig, amssymb, color, bm}
\usepackage{latexsym}
\usepackage{graphicx}
\usepackage{accents}
\usepackage{subfigure}
\usepackage{epsfig}
\usepackage{color}

\begin{document}

\title{Check Reliability Based Bit-Flipping Decoding Algorithms for LDPC Codes}

\author{\authorblockN{Chi-Yuan Chang, Yu T. Su, Yu-Liang Chen, and Yin-Chen Liu}
\authorblockA{Institute of Communications Engineering \\National Chiao Tung University\\
Hsinchu, 30056, TAIWAN,\\ Email: ~ ytsu@mail.nctu.edu.tw}}

\maketitle

\linespread{1.03}
\begin{abstract}
We introduce new reliability definitions for bit and check nodes.
Maximizing global reliability, which is the sum reliability of all
bit nodes, is shown to be equivalent to minimizing a decoding
metric which is closely related to the maximum likelihood decoding
metric. We then propose novel bit-flipping (BF) decoding
algorithms that take into account the check node reliability. Both
hard-decision (HD) and soft-decision (SD) versions are considered.
The former performs better than the conventional BF algorithm and,
in most cases, suffers less than 1 dB performance loss when
compared with some well known SD BF decoders. For one particular
code it even outperforms those SD BF decoders. The performance of
the SD version is superior to that of SD BF decoders and is
comparable to or even better than that of the sum-product
algorithm (SPA). The latter is achieved with a complexity much
less than that required by the SPA.
\end{abstract}

\IEEEpeerreviewmaketitle

\section{Introduction}
Low-density parity-check (LDPC) codes were first introduced by
Gallager \cite{LDPC} in early 1960s and rediscovered by Mackay
\cite{NSLDPC} \cite{GEC} in 1990s. Soft-decision, hard-decision
and hybrid decoding algorithms have been proposed for decoding
LDPC codes. The sum-product algorithm (SPA) achieves the
near-capacity performance asymptotically but its computational
complexity is very high. The min-sum algorithm (MSA)
\cite{RCLDPC}, which replaces the nonlinear check node operation
by a single minimum operation, was proposed to reduce the
complexity of the standard SPA at the cost of a noticeable
degradation in the decoding performance. Chen and Fossorier
\cite{DEBP}-\cite{RCDLDPC} suggested the normalized min-sum
algorithm and the offset min-sum algorithm which multiply and add
a constant correction factor in the check-to-variable updating
equation of MSA. They offer performance compatible to that of the
SPA but with lower complexity.

Bit-flipping (BF) decoding algorithm is a hard-decision decoding
algorithm which is much simpler than SPA or its modifications but
does not perform as well. To reduce the performance gap between
SPA and BF based decoders, variants of the latter such as weighted
bit-flipping (WBF) \cite{WBF}, modified weighted bit-flipping
(MWBF) \cite{MWBF} and improved modified bit-blipping (IMWBF)
\cite{IMWBF} algorithms have been proposed. They provide tradeoffs
between computational complexity and error performance. The
reliability ratio based weighted bit-flipping (RRWBF) \cite{RRWBF}
decoding algorithm needs not to find optimal parameters as
variants of the WBF algorithm do but yields better performance.

In this paper, we present novel bit-flipping algorithms called
check reliability based bit-flipping (CRBF) decoding algorithms
for decoding LDPC codes. Starting with the maximum likelihood (ML)
decoding metric, we first relax the parity-check requirements and
introduce a {\it global cost} which is the sum $N$ local costs,
where $N$ is the codeword length. Interpreting each cost as the
opposite of reliability, the bit(s) with maximal local cost or
least reliability is (are) flipped in each decoding iteration. By
reducing the local costs iteratively, we hope that the global cost
can also be minimized to approach the ML cost. We define the
reliability of a check node and use this reliability to modify and
update the bit (node) reliabilities.

Two CRBF algorithms are proposed: the soft check reliability based
bit-flipping (soft-CRBF) algorithm, which processes the received
channel values when decoding, and its hard decision counterpart
which sends the hard-decision demodulated bit streams to the
decoder. The soft-CRBF outperforms the WBF decoding algorithm and
its variants and is comparable to SPA for some LDPC codes. We also
compare the performance of the hard-CRBF and standard BF decoders
and the simulation results prove that the former is a better
choice.

The rest of this paper is organized as follows. In Section
\ref{subsection:Cost}, we introduce the global and local costs as
minimizing decoding costs and derive the relation between the
global cost and the ML cost. The check node cost and reliability
are defined in Section \ref{subsection:Reliability}. In Section
\ref{subsection:CRBF}, we describe the proposed CRBF algorithms,
and in \ref{subsection:WBFs} the cost functions of conventional
WBF algorithms are examined. Section \ref{section:complexity}
discusses the computational complexities of the proposed
algorithms, and Section \ref{section:Results} provides some
simulated numerical results concerning the performance of our and
some well known algorithms. Finally, conclusion remarks are drawn
in Section \ref{section:conclusion}.

\section{Check Reliability Based Bit-Flipping Decoding Algorithms}
\subsection{Cost Functions}\label{subsection:Cost}
A regular binary ($N$, $K$) ($d_v$, $d_c$) LDPC code $\mathcal{C}$
is a linear block code described by an $M \times N$ parity check
matrix ${\textbf{H}}$ which has constant column weight of $d_v$
and row weight of $d_c$. ${\textbf{H}}$ can be represented by a
bipartite graph with $N$ variable nodes corresponding to the
encoded bits, and $M$ check nodes corresponding to the
parity-check functions represented by the rows of ${\textbf{H}}$.
The code rate of $\mathcal{C}$ is given by $\mathcal{R}= K/N$.

Assume BPSK signaling with unit energy is used. A codeword
$\bm{c}=(c_0, c_1,\dots,c_{N-1})$ is mapped into a bipolar
sequence $\hat{\bm{c}}=(\hat{c}_0,\hat{c}_1,\dots,\hat{c}_{N-1})$
and transmitted over an AWGN channel with noise variance
$\sigma^2$. Let $\bm{y}=(y_0,y_1,...,y_{N-1})$ be the
corresponding received soft-decision sequence and the binary
hard-decision sequence, $z=(z_0,z_1,...,z_{N-1})$, is obtained as
follows:
\begin{eqnarray}
z_i=\left\{\begin{array}{ll} 0&, \text{if } y_i\geq 0\\1&,
\text{if } y_i<0
\end{array}
\right..
\end{eqnarray}
Let $\bm{x}^{l}=(x_0^l, x_1^l,\dots,x_{N-1}^l)$ be the decoded
sequence at the $l$th iteration, and $\bm{s}^{l}=(s_0^l,
s_1^l,\dots,s_{M-1}^l)$ be the syndrome vector of $\bm{x}^l$:
\begin{eqnarray}
\bm{s}^l = (s_0^l, s_1^l,\dots,s_{M-1}^l) =
\bm{x}^l\textbf{H}^\text{T}.
\end{eqnarray}
Define $\hat{\bm{z}}=(\hat{z}_0, \hat{z}_1,\dots,\hat{z}_{N-1})$,
$\hat{\bm{x}}^{l}=(\hat{x}_0^l,
\hat{x}_1^l,\dots,\hat{x}_{N-1}^l)$ and $\hat{\bm{s}}^l =
(\hat{s}_0^l, \hat{s}_1^l,\dots,\hat{s}_{M-1}^l)$ be the bipolar
modulated sequences corresponding to $\bm{z}$, $\bm{x}^l$ and
$\bm{s}^l$, respectively. We thus have $\hat{z}_i=1-2\cdot z_i$,
$\hat{x}_i^l=1-2\cdot x_i^l$ for $0\leq i<N$,
$\hat{s}_i^l=1-2\cdot s_i^l$ for $0\leq i<M$, and
$\hat{\bm{z}}^{l}, \hat{\bm{x}}^{l} \in \{1,
-1\}^N\stackrel{def}{=}F_2^N$.

Let $\mathcal{N}$($m$) be the set of variable nodes that
participate in check node $m$ and $\mathcal{M}$($n$) be the set of
check nodes that are connected to variable node $n$ in the code
graph. $\mathcal{N}(m)\backslash n$ is defined as the set
$\mathcal{N}$($m$) with the variable node $n$ excluded while
$\mathcal{M}(n)\backslash m$ is the set $\mathcal{M}$($n$) with
the check node $m$ excluded.

Maximum likelihood (ML) decoding would find the unique bipolar
vector
$\tilde{\bm{x}}=\arg\max_{\hat{\bm{x}}\in\mathcal{C}}\sum_{i=0}^{N-1}\hat{x}_iy_i$,
where the constraint ${\hat{\bm{x}}\in\mathcal{C}}$ is equivalent
to $\hat{s}_j=\prod_{j^{'}\in\mathcal{N}(j)}\hat{x}_{j^{'}},
~\forall~0\leq j \leq M-1$. Equivalently, the ML decoder solves
the unstrained optimization problem
\begin{equation}
\arg\min_{\hat{\bm{x}} \in F_2^N}
-\left[\sum_{i=0}^{N-1}\hat{x}_iy_i+\sum_{j=0}^{M-1}\alpha_j(\hat{s}_j-1)\right]
\label{MLCF}
\end{equation}
where $\alpha_i$ are Lagrange multipliers. Although $\alpha_i$'s
can be arbitrary real numbers the fact that each $\hat{s}_j$ can
only take two values, +1 and -1, implies that they must be
positive for otherwise the cost function will encourage the
violation of the constraints $\hat{s}_j=1$. Instead of solving the
above ML problem we attack a simpler problem by relaxing the
constraints and define a \emph{global cost} (GC) associated with a
candidate bipolar $N$-tuple $\hat{\bm{x}}$ as
\begin{eqnarray}
E(\hat{\bm{x}})\triangleq-\sum_{i=0}^{N-1}\hat{x}_iy_i-\alpha\sum_{j=0}^{M-1}\hat{s}_j,\label{GlobalCF}
\end{eqnarray}
where $\alpha > 0$. The above cost replaces the multiple
constraints $\{\hat{s}_i-1=0, i=0,1,\cdots, M-1\}$ by the single
constraint $\sum_{i=0}^{M-1}\hat{s}_i=M$. The second term in the
above equation is used to penalizes each invalidate check relation
and the penalty is minimized if $\hat{\bm{x}}$ is a valid
codeword. Unlike (\ref{MLCF}) which penalizes each violation
$\hat{s}_j\neq 1$ differently, (\ref{GlobalCF}) imposes a constant
penalty.

(\ref{GlobalCF}) can be rewritten as
\begin{eqnarray}
E(\hat{\bm{x}})&=&
-\sum^{N-1}_{i=0}\left(\hat{x}_iy_i+\frac{\alpha}{d_c}\sum_{j\in\mathcal{M}(i)}\hat{s}_j\right)\nonumber\\
&=&-\sum^{N-1}_{i=0}\left(\hat{x}_iy_i+\gamma\sum_{j\in\mathcal{M}(i)}\hat{s}_j\right)\nonumber\\
&=&\sum^{N-1}_{i=0}E_i,
\end{eqnarray}
where $\gamma=\frac{\alpha}{d_c}$ is also a positive constant. We
accordingly define the \emph{local cost} (LC) for variable node
$i$ by
\begin{eqnarray}
E_i\triangleq
-\left(\hat{x}_iy_i+\gamma\sum_{j\in\mathcal{M}(i)}\hat{s}_j\right)
\end{eqnarray}
and interpret the quantity $-E_i$, if positive, as the reliability
of variable node $i$. Minimizing the global cost is thus equal to
maximizing the total (variable node) reliability. Note that the
LC's are not independent and related through $\hat{s}_i$ unless
$\mathcal{M}(i) \cap \mathcal{M}(j) = \emptyset, \forall~ j \neq
i$.
\subsection{Reliability of a Check Node}\label{subsection:Reliability}
Given the reliability and local cost of a variable node, we now
define the corresponding reliability and local cost of a check
node by
\begin{eqnarray}
R_{mn}= \max ( -R^*_{mn}, 0) \label{RBofC}
\end{eqnarray}
where
\begin{eqnarray}
R^{*}_{mn}=\max_{n^{'}\in\mathcal{N}(m)\backslash n}E_{n^{'}}
\label{MaxCost}
\end{eqnarray}
$R^{*}_{mn}$ is the maximum local cost amongst those of the
variable nodes connecting to the check node $m$ except node $n$.
In other words, $R^{*}_{mn}$ is used as a measure of the
\emph{unreliability} of the massage check node $m$ intends to pass
to variable node $n$. This unreliability is equal to the maximal
cost of the variable nodes in $\mathcal{N}(m)\backslash n$. If the
maximum local cost (unreliability) is too high, (\ref{RBofC}) will
return a constant zero, which means the check node $m$ is totally
useless for variable node $n$. In contrast, the reliability
$R_{mn}$ is a positive weighting factor which indicates whether
the check node (relation) $m$ can provide proper reliability
information for variable node $n$. Note that a bit-flipping on
variable node $i$ will result in a magnitude change of $E_i$. The
maximum magnitude change is $|y_i|+\gamma |\mathcal{M}(i)|$, where
$|\mathcal{M}(i)|$ denotes the cardinality of $\mathcal{M}(i)$.
The definition of the check reliability in (7) and (8) indicates
this maximum magnitude change is also the maximum candidate value
for $R_{mn}$ which occurs when a incorrect bit decision is made
and all check relations are violated.
\subsection{Check Reliability Based Bit-Flipping Decoding Algorithms}\label{subsection:CRBF}
When using an iterative algorithm to find the minimum GC,
$\hat{x}_i$ and $\hat{s}_i$ are replaced by their $l$th iteration
values $\hat{x}^l_i, \hat{s}^l_i$ in computing (\ref{GlobalCF}).
To use the check reliability information we update the LC at the
$l$th iteration by
\begin{eqnarray}
E_i^{l}\stackrel{def}{=}
-\left(\hat{x}^{l}_iy_i+\gamma\sum_{j\in\mathcal{M}(i)}R^l_{ji}\hat{s}_j^l\right),
\label{newE}
\end{eqnarray}
where
\begin{eqnarray}
R_{ji}^{l}=\max (-R^{*l}_{ji}, 0)\label{mRBofC}
\end{eqnarray}
is the reliability of check node $j$ for variable node $i$ in the
$l$th iteration and
\begin{eqnarray}
R^{*l}_{ji}=\max_{i^{'}\in\mathcal{N}(j)\backslash
i}E^{l-1}_{i^{'}}-\gamma
\hat{s}_j^{l-1}R^{l-1}_{ji^{'}}.\label{muRBofC}
\end{eqnarray}
is the \emph{modified unreliability} of the check node $j$ for
variable node $i$. Since $\gamma > 0$, (\ref{newE}) and
(\ref{muRBofC}) indicate that we have put larger weights on more
reliable checks.

The basic procedure of the proposed CRBF decoding algorithms works
as follows. At each iteration we flip the decision bit(s) which is
(are) most unreliable (largest LC) first, calculate the
reliability of each check node, and update the LC of each variable
node if needed. The procedure stops if a validate codeword is
found or if the maximum number of iterations $I_{max}$ is reached.
The \emph{soft} CRBF (soft-CRBF) decoding algorithm is summarized
in Table I.

\begin{table}
\vspace{2ex} \caption{\label{Algo}A Soft Check Reliability Based
Bit-Flipping Decoding Algorithm} \normalsize{}
\begin{tabular}{l}
\hline \hline

\noindent$\textbf{Initialization:}$\\
Set $l=0$, $\hat{\bm{x}}^{0}\leftarrow \hat{\bm{z}}$, compute
$\hat{\bm{s}}^0$.\\Let $R^0_{ji}=1$, $\forall~ i\in\mathcal{N}(j)$, $0\leq j<M$.\\
$E_i^0=-y_i-\gamma\sum_{j\in\mathcal{M}(i)}\hat{s}_j^0$, for
$0\leq i<N$. \\

\noindent$\textbf{Step 1:}$\\
$l=l+1$. $\hat{x}^l_i=\hat{x}^{l-1}_i$, $0\leq i<N$.\\
$e=\arg\max_{i}E_i^{l-1}$, then let
$\hat{x}^l_e=-\hat{x}^{l-1}_e$.
\\
\noindent$\textbf{Step 2:}$\\
Compute $\bm{s}^{l}$. If $\bm{s}^{l}=\bm{0}$ or $l=I_{max}$, stop decoding \\
and output $\bm{x}^{l}$ as the decoded sequence. Otherwise, \\go
to \textbf{step 3}.
\\
\noindent$\textbf{Step 3:}$\\
$\forall i\in\mathcal{N}(j)$ and $0\leq j<M$, compute $R_{ji}^l$.
\\
\noindent$\textbf{Step 4:}$\\
For $0\leq i<N$, compute $E_i^l$ and
go to \textbf{step 1}.

\\\hline
\hline
\end{tabular}
\end{table}

If we replace $y_i$ with $\hat{z}_i$ in the $\textbf{Step 4}$ of
Table I, the resulting algorithm is called the \emph{hard} CRBF
(hard-CRBF) decoding algorithm.
\subsection{Cost Functions of Known WBF Decoding Algorithms}\label{subsection:WBFs}
\label{subsection:complex} For the weighted bit-flipping (WBF)
decoding, the cost function or the objective function for
$\hat{x}_i^l$ in the $l$th iteration is defined by
\begin{eqnarray}
E_{i,WBF}^l\triangleq -\sum_{j\in\mathcal{M}(i)}\hat{s}_j^l\cdot
w_{ji},
\end{eqnarray}
where
\begin{eqnarray}
w_{j,i}=\min_{i^{'}\in\mathcal{N}(j)\setminus i}|y_{i^{'}}|.
\end{eqnarray}
For modified WBF (MWBF) and improved modified WBF (IMWBF) decoding
algorithms, the cost functions for $\hat{x}_i^l$ in the $l$th
iteration is defined by
\begin{eqnarray}
E_{i,MWBF}^l\triangleq -\sum_{j\in\mathcal{M}(i)}\hat{s}_j^l\cdot
w_{ji}-|y_i|
\end{eqnarray}
and
\begin{eqnarray}
E_{i,IMWBF}^l\triangleq -\sum_{j\in\mathcal{M}(i)}\hat{s}_j^l\cdot
w_{ji}-\alpha|y_i|
\end{eqnarray}
respectively, where $\alpha$ is a positive constant and can be
optimized by simulations. The bit with maximal cost will be
flipped in the WBF, MWBF and IMWBF decoding algorithms.

\section{Computational Complexity} \label{section:complexity}
For the proposed algorithm, each decoding iteration consists of
check-reliability and the cost updates.
The number of check-reliability updates in the proposed algorithms
is highly dependent on the node degrees of the code used. Suppose
only one bit is flipped in each iteration, we will need
$\min\{M+1,d_v\cdot(d_c-1)+1\}$ standard operations, each includes
the check-reliability updates and the selection of the bit with
the maximum local cost, in one iteration. $\min\{N,d_v\cdot d_c\}$
cost calculations are also needed.

For the WBF, MWBF and IMWBF decoding algorithms, $w_{j,i}$ is used
to adjust the reliability of check node $j$ for variable node $i$.
$w_{j,i}$ will not be modified in the course of iterative
decoding, so the computational complexity is lower than that of
the proposed algorithms. The standard BF algorithm can be regarded
as a special case of the WBF decoder with $w_{j,i}=1$, which has
the lowest complexity but gives relatively poor performance.

\section{Numerical Results}\label{section:Results}
Computer-simulated bit error rate (BER) performance of various BF
decoding methods, SPA and proposed decoding algorithms are
reported in this section. Mackay's (504,252)(3,6) 0.5-rate LDPC
code, (255,175)(16,16) 0.69-rate EG-LDPC code, the
(1440,1344)(3,45) 0.93-rate LDPC code defined in 802.15.3c and the
(2048,1723)(6,32) 0.84-rate LDPC code defined in 802.3a/n are used
for comparison in our simulations.

\begin{figure}
\begin{center}
\epsfxsize=3.5in \epsffile{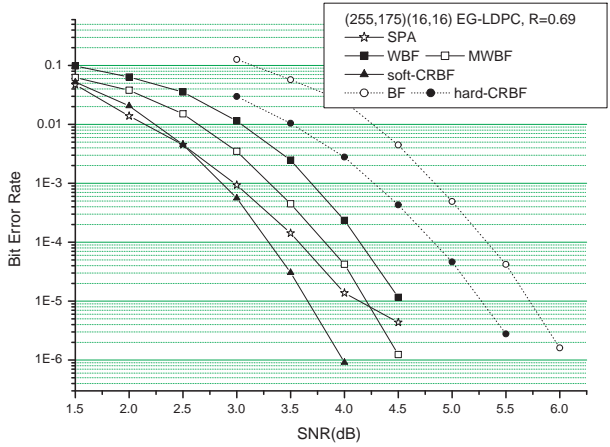}\caption{\label{BER255} BER
performance of the (255,175)(16,16) LDPC code using different
decoding methods: SPA, BF, WBF, MWBF, IMWBF, hard-CRBF and
soft-CRBF algorithms.}
\end{center}
\end{figure}
\begin{figure}
\begin{center}
\epsfxsize=3.5in \epsffile{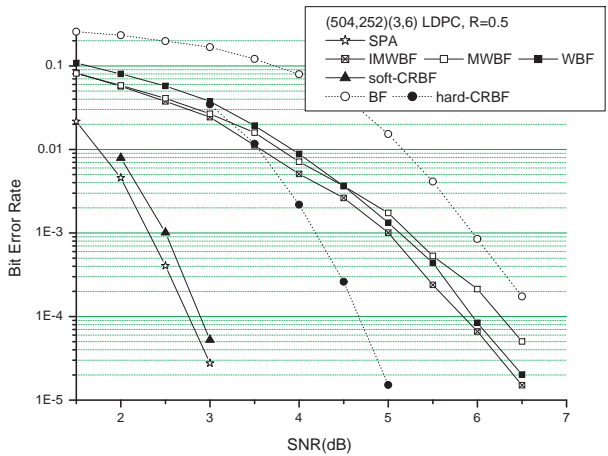}\caption{\label{BER504} BER
performance of the SPA, BF, WBF, MWBF, IMWBF, hard-CRBF and
soft-CRBF decoding algorithms for the (504,252)(3,6) LDPC code.}
\end{center}
\end{figure}
\begin{figure}
\begin{center}
\epsfxsize=3.5in \epsffile{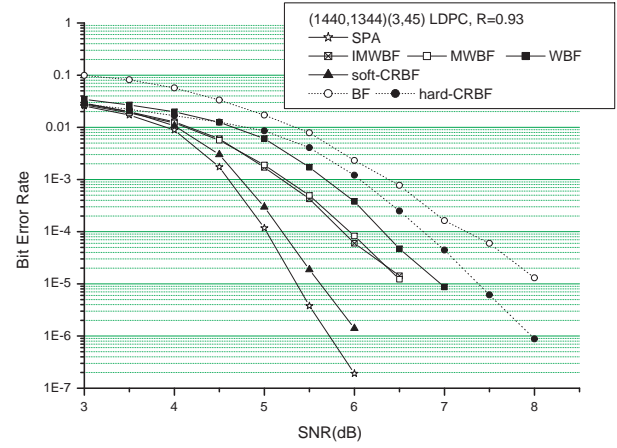}\caption{\label{BER1440}
BER performance of the (1440,1344)(3,45) LDPC code using the SPA,
BF, WBF, MWBF, IMWBF, hard-CRBF and soft-CRBF algorithms.}
\end{center}
\end{figure}
\begin{figure}
\begin{center}
\epsfxsize=3.5in \epsffile{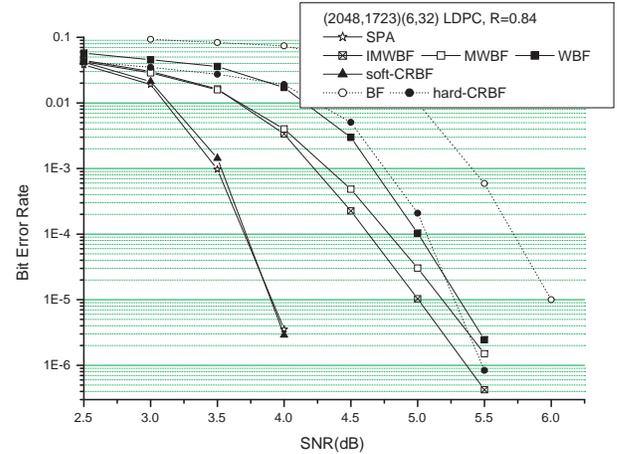}\caption{\label{BER2048}
BER performance of the (2048,1723)(6,32) LDPC code with the SPA,
BF, WBF, MWBF, IMWBF, hard-CRBF and soft-CRBF algorithms.}
\end{center}
\end{figure}
Fig.\ref{BER255} depicts the BER performance of the standard BF,
WBF, MWBF, soft-CRBF and hard-CRBF decoding algorithms with
$I_{max}=30$. The performance of IMWBF decoding is not shown in
this figure because the optimal parameter value for $\alpha$
equals to 1, that is, the two algorithms, IMWBF and MWBF, become
identical for this code. The proposed soft-CRBF algorithm
outperforms the SPA, WBF and MWBF by about 0.35 dB, 0.5 dB, and
0.8 dB, respectively at BER$=10^{-5}$. The hard-CRBF decoding
algorithm outperforms the standard decoding algorithm, which is
also a pure hard-decision decoding algorithm, by about 0.5 dB at
BER$=4\times10^{-4}$.

Performance curves in Fig. \ref{BER504} show that at
BER$\approx10^{-4}$, the proposed soft-CRBF algorithm gives 3 dB
perform gain with respect to the variants of WBF algorithms. The
hard-CRBF algorithm offers 2 dB perform gain against the standard
BF decoder at BER$=2\times10^{-4}$ and provides better performance
than the variants of WBF algorithms; $I_{max}$ being set as $70$
for this code.

Fig. \ref{BER1440} compares the BER performance of the SPA,
variants of the WBF algorithm, and the proposed algorithms with
$I_{max}=30$. The soft-CRBF decoding algorithm outperforms the
IMWBF and the MWBF decoders by about  0.9 dB at BER$=10^{-5}$ and
yields 1.3 dB gain against the WBF decoder at the same BER. The
hard-CRBF decoding algorithm is about 0.6 dB better than the
standard BF at BER$=1.2\times10^{-5}$.

Fig. \ref{BER2048} indicates that the proposed soft-CRBF decoder
yields near-SPA performance and outperform the WBF, MWBF and IMWBF
decoder by a margin larger than 1 dB at BER$=10^{-5}$. The
hard-CRBF not only outperforms the standard BF but is also
superior to the MWBF and IMWBF algorithms at BER$=10^{-6}$ when
$I_{max}=70$.

\section{Conclusion}\label{section:conclusion}

We have presented two novel check reliability based soft-decision
bit-flipping decoding algorithms to improve the performance of the
WBF algorithm and its variants for decoding LDPC codes. At each
iteration, the cost/reliability for each bit is computed and the
bit with least reliability is flipped. The check reliability is
also defined for each check node and is used to update the related
bit node reliabilities. The sum of bit cost/reliability is shown
to be a relaxed version of the ML decoding metric. Our algorithms
are iterative approaches for minimizing the sum reliability.
Numerical results show that the proposed soft-decision decoding
algorithm outperforms the conventional WBF algorithm and its
variants. On the other hand, the hard-decision version outperforms
the standard bit-flipping decoder and, for some codes, even offers
performance better than that of the WBF decoding algorithm.
\begin{center}
ACKNOWLEDGEMENT
\end{center}
This work was supported in part by Taiwan's National Science
Council under Grant NSC-96-2221-E-009-076-MY3.

\maketitle 

\end{document}